\documentclass[reprint,bibnotes,amsmath,amssymb,aps,prb,showpacs,floatfix,superscriptaddress,bibliography]{revtex4-1}
\usepackage[utf8]{inputenc}
\usepackage{amsmath}
\usepackage{mathtools}
\usepackage{bm}
\usepackage{amsfonts}
\usepackage{amssymb}
\usepackage[breaklinks=true,colorlinks,citecolor=blue,linkcolor=blue,urlcolor=blue]{hyperref}
\usepackage{graphicx}
\newcommand{\RN}[1]{%
	\textup{\uppercase\expandafter{\romannumeral#1}}%
}

\def\be{\begin{equation}}
\def\ee{\end{equation}}
\def \bea{\begin{eqnarray}}
\def \eea{\end{eqnarray}}
\def \nn{\nonumber}
\def \moire{moir\'e }
\usepackage{xcolor}
\renewcommand{\thefootnote}{\alph{footnote}}
\usepackage[symbol]{footmisc}
\renewcommand{\thefootnote}{\fnsymbol{footnote}}

\begin{document}
	\title{Tunable interband and intraband plasmons in twisted double bilayer graphene}
	\author{Atasi Chakraborty$^\parallel$}
	\email{atasic@iitk.ac.in}
   	\affiliation{Department of Physics, Indian Institute of Technology Kanpur, Kanpur-208016, India}
	\author{Debasis Dutta$^\parallel$}
	\email{ddebasis@iitk.ac.in}
	\affiliation{Department of Physics, Indian Institute of Technology Kanpur, Kanpur-208016, India}
	\author{Amit Agarwal}
	\email{amitag@iitk.ac.in}
	\thanks{\\\noindent$\parallel$ These authors contribute equally to this work.}
	\affiliation{Department of Physics, Indian Institute of Technology Kanpur, Kanpur-208016, India}

\begin{abstract}
Flat bands in twisted \moire  superlattices support a variety of topological and strongly correlated phenomena along with easily tunable electrical and optical properties. 
Here, we demonstrate the existence of tunable, long lived, and flat intraband and interband terahertz plasmons in twisted double bilayer graphene. 
We show that the interband plasmons originate from the presence of a Van Hove singularity in the joint density of states and a finite Berry connection between the pair of bands involved. We find that the gapped interband plasmon mode has a universal dispersion, 
and the plasmon gap is specified by the location of the Van Hove singularity in the joint density of states. Metallic \moire systems support an additional intraband plasmon mode which becomes flat in the 
large momentum limit because of the influence of the interband correlations. % of the flat electronic bands.
We demonstrate that the undamped and flat plasmon modes in \moire systems are highly tunable, and can be controlled by varying the vertical electric field, electron doping, and they persist over a wide range of twist angles. 
% Our study elucidates the tunable interband plasmons on \moire systems for future realization of nano-photonic applications.
\end{abstract}
\maketitle 
\def\thefootnote{*}\footnotetext{These authors contributed equally to this work}

\section{Introduction}\label{{section-I}}
\noindent Plasmons are the self-sustained  collective modes of electronic charge density oscillations in materials hosting electron liquids~\cite{Pines1952, Pines1962, Pines1966}.
Plasmons act as bridges for efficient light-matter interactions which is essential for harnessing the combined power of the fast optical timescales and the small lattice length-scales. Dissipation-free plasmon modes form an important component of the light-based quantum computing toolbox~\cite{Quantum_computing2019,Tame2013}. Long lived and tunable plasmon modes are crucial for applications based on quantum plasmonics, dissipationless light-matter interactions and nano-photonics applications~\cite{BingHuang2021,Politano_t_2018,Rivera2020,Koppens2013,Song2015,Sadhukhan_N_2020,Yu2019}. In particular, plasmons in two-dimensional (2D) materials have immense potential for opto-electronic applications owing to their long lifetime ($\sim 500$~fs)~\cite{Yan2013, Andress2012, Agarwal_L_2014, Agarwal2018,Wang2021_applications}, large propagation distance~\cite{Marco2015}, electrostatic tunability and sub wavelength confinement~\cite{Gao2012, Zhou2012,Jablan2009,Fei2012} over a broad spectral range. 2D materials such as graphene generally host two kind of plasmons. The gapless intraband plasmon arising from resonant density fluctuations around the Fermi surface, and the interband plasmon arising from resonant interband density fluctuations~\cite{Sachdeva_P_2015, Jablan2009, SDSharma2007, Kinyanjui_2012, Despoja2013, Antonio2014}. However, both of these modes in graphene are dispersing and 
get damped beyond a certain momentum due to their proximity to single-particle electron hole excitations.  

The recent discovery of small angle \moire superlattices with unique  electronic properties have opened up new  directions for exploring tunable opto-electronic properties in 2D systems~\cite{Basov2015,Wang2020_application}. In \moire systems, the small twist angle ($\sim 1^\circ$) induced spatial variation of the interlayer couplings suppress the dispersion of the electronic states and gives rise to topological flat bands \cite{Bistritzer2011, Fu2020,Haddadi2020}. The flat electronic bands give rise to Van Hove singularities (VHS) in the density of states (DOS), which makes these a playground for many strongly correlated phenomena, such as ferromagnetism and superconductivity, among others~\cite{burg_correlated_2019, MacDonald2020, shen2020, cao_tunable_2020, Sinha_B_2022,Guinea2021,Liu2020,Bennett2022}. Moir\'e superlattices such as twisted bilayer graphene (TBG) have been shown to support undamped and long lived intraband and interband plasmon modes with a flat dispersion ~\cite{marco_tbg20,Stauber2016,Levitov2019,Shengjun2021,Basov2015}. In fact, an interband plasmon mode in bernal-stacked TBG was recently demonstrated through mid-infrared near-field optical microscopy \cite{FrankHLKoppen2021}. The flat plasmon modes in \moire superlattices can also possibly mediate unconventional superconductivity~\cite{Girish2020, Cyprian2021}.  However, the physics of the interband plasmon modes,  the criteria for their existence, and the origin of the flat plasmon dispersion in the interband as well as in the intraband plasmon mode is not clearly understood. 

Here, we demonstrate the existence of a ladder of long lived, flat, and gate tunable intraband and interband plasmons in the  \moire superlattice of twisted double bilayer graphene (TDBG). We establish the universality of the gapped interband plasmon modes in systems with i) a VHS in the joint density of states (JDOS), and ii) a finite interband Berry connection. We show that the interband plasmon modes in 2D have a universal long wavelength  dispersion of the form, 
\be \omega_{\rm inter} (q \to 0) = \Delta_0 \sqrt{1 + \gamma ~q}~.\ee 
Here, $\Delta_0$ marks the location of the VHS peak in the JDOS, and $\gamma$ is a material specific parameter. Similarly to the flat electronic bands, the interband plasmon dispersion in TDBG also becomes flat in the large $q$ limit. We find that in addition to the interband plasmon mode, metallic TDBG also host a gapless intraband plasmon mode which disperses universally as $\sqrt q$ in the long wavelength limit. However, it becomes flat in the large $q$ limit owing to the interband screening effects. We demonstrate that this ladder of interband and intraband plasmon modes is highly tunable and can be controlled by varying the twist angle, electronic doping, and by an externally applied vertical electric field.
\begin{figure*}[t!]
 	\centering
 	\includegraphics[width=\linewidth]{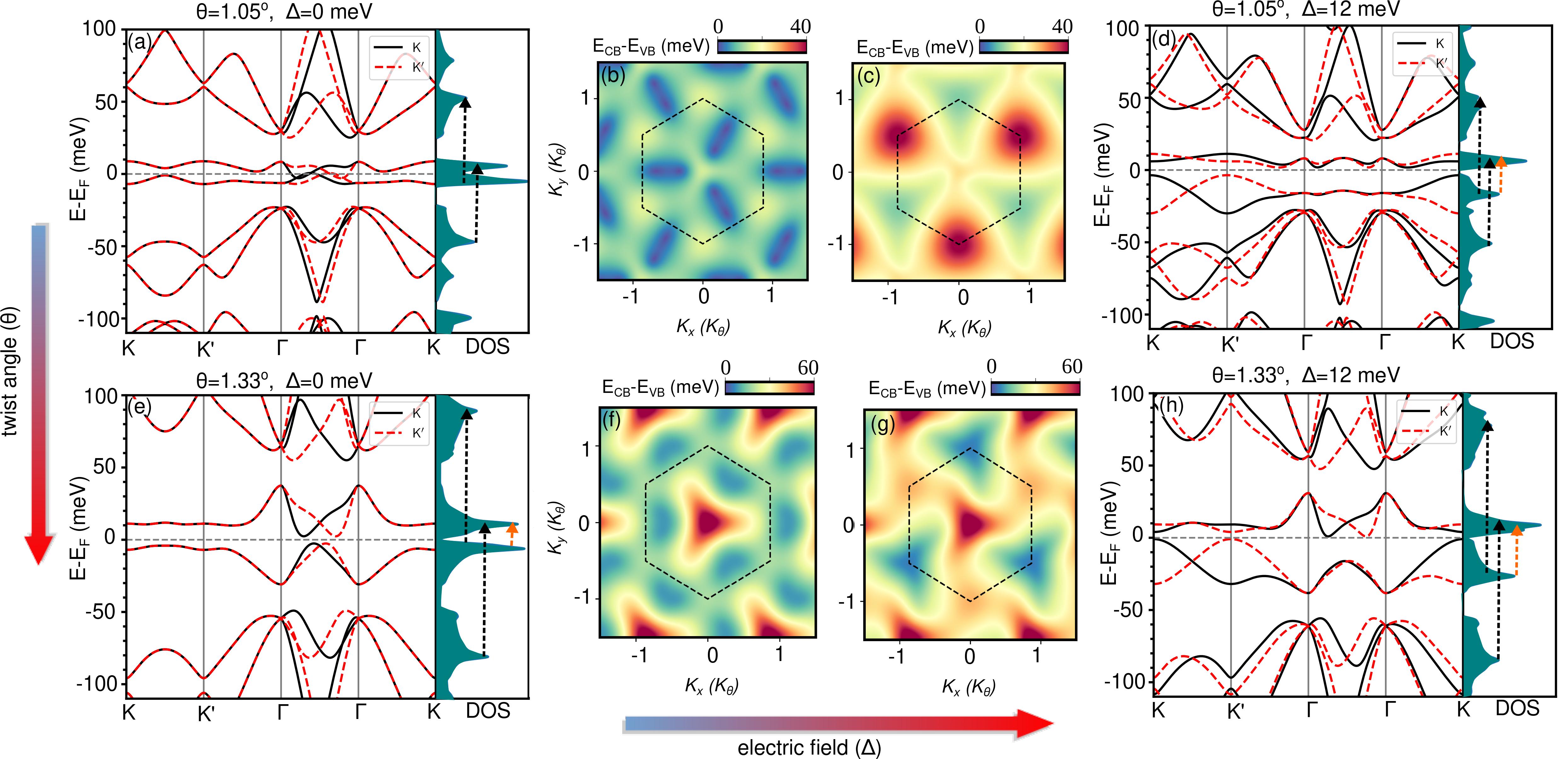}
 	\caption{(a), (d) and (e), (h) show the electronic band dispersion of TDBG for two different twist angles $\theta=1.05^{\circ}$ and $1.33^{\circ}$ in absence ($\Delta=0$~meV) and in presence of finite ($\Delta=12$~meV) perpendicular electric field respectively. The black solid and red dotted lines represent the band dispersion for the $K$ and $K^\prime$ valley respectively. (b), (c), (f) and (g) display the corresponding direct band gap [$E_{\rm CB}({\bf k}) -E_{\rm VB}({\bf k}) $] between the first conduction and first valence bands within the whole Brillouin zone. Clearly the electronic band structure is highly tunable on varying the perpendicular electric field, and the twist angle.}
 	\label{fig1}
 \end{figure*}
The gate-tunable, long lived, and slow plasmon modes in \moire superlattices offer an intriguing platform for exploring    opto-electronic applications.   

%%%%%%%%%%%%%%%%%%%%%%%%%%%%%%%%%%%%%%%%%%%%%%%%%%%%%%%%%%%%%%%%
%\section{Results}
\section{Moir\'e flat bands and Van Hove Singularity}\label{section-II}
\noindent To study plasmons in \moire superlattice of TDBG, we start with 
 its electronic band structure and its tunability. 
At a small twist angle, the electronic spectrum  of TDBG can be described through the low energy extended continuum model Hamiltonian, following the approach originally proposed by \textit{Bistritzer-MacDonald (BM)}~\cite{Bistritzer2011, Andrei2020} for TBG. The details of the continuum model calculations are presented in Sec.~A1 of the Supporting information (SI)~\cite{Note1}. We find that for small values of the twist angles, there is a pair of flat bands in the vicinity of the charge neutrality point (CNP), well separated from the other higher \moire bands. The `flatness' of these low energy bands give rise to VHS in the DOS. The presence of VHS in \moire superlattices makes these systems susceptible to correlation effects, leading to several exotic phases ~\cite{burg_correlated_2019, MacDonald2020, shen2020, cao_tunable_2020, Sinha_B_2022}. 

More interestingly, the band dispersion of twisted \moire superlattices, in vicinity of the Fermi energy, are very sensitive to the twist angle ($\theta$), and to the external electric field ($\Delta$) applied perpendicular to the 2D lattice plane. Both these parameters serve as experimental knobs for manipulating the electronic properties of \moire superlattices \cite{Choi2019,Sinha2020,Sinha_B_2022, shen2020}. As an example, we show the evolution of the flat bands of the AB-AB stacked TDBG with the variation of the twist angle and externally applied vertical electric field in Fig.~\ref{fig1}. Fig.~\ref{fig1} (a) shows the band dispersion along the high symmetry directions in the Brillouin zone (BZ), together with the DOS at the \textit{magic} angle, $\theta= 1.05^\circ$. Similar to the case of TBG, we use the word `\textit{magic} angle' to imply that the bandwidths of the first conduction and the first valence bands are the smallest possible for this specific angle.  Clearly, at $\theta=1.05^\circ$ the first conduction and first valence bands are almost completely flat (bandwidth within $\sim10$ meV range) throughout the whole BZ (see Fig.~\ref{fig1} (b)). The overlapping flat bands induce finite electronic states at the Fermi energy, as reflected in the DOS plot in the right panel of Fig.~\ref{fig1}(a). The low energy flat bands are separated from the higher bands with a finite energy gap, referred to as the \moire gap, of value $\sim 16$~meV. The increment in $\theta$ gradually increases the bandwidth of the flat bands, in addition to inducing a finite energy gap between the two flat bands in vicinity of the CNP [see Fig.~\ref{fig1} (e), (f)]. The band structure and DOS for TDBG at a slightly higher twist angle $\theta= 1.33^\circ$ are presented in Fig.~\ref{fig1} (e). Although the pair of flat bands near the Fermi energy are relatively more dispersing around the $\Gamma$ point, the flat characteristics of the bands persist over rest of the BZ (see Fig.~\ref{fig1} (f)). This also appears as VHS in the corresponding DOS. The three-fold rotation ($C_3$) symmetry of TDBG is clearly depicted in the direct band gap plots of Fig.~\ref{fig1} (b) and (f). The absence of any zero value in Fig.~\ref{fig1} (f), establishes the insulating nature of the ground state of TDBG with $\theta = 1.33^\circ$.

Compared to TBG, TDBG offers a higher tunability of its electronic properties with an external electric field. The impact of electric field is introduced in the Hamiltonian via a diagonal potential term ($\Delta$) as presented in detail in the Sec.~A1 of SI~\cite{Note1}.  Its impact on electronic properties is represented by the horizontal arrow in Fig.~\ref{fig1}.  The band dispersion together with DOS of TDBG at $\theta=1.05^\circ$ and $1.33^\circ$ in presence of $\Delta=12$~meV are presented in Fig.~\ref{fig1} (d) and (h), respectively. At the \textit{magic} angle, the pair of narrow bands at the CNP start to gap out on application of finite electric field beyond a critical threshold. We calculate the critical field required to promote an overall insulating state at the \textit{magic} angle to be $\sim 9$~meV. The DOS plot of Fig.~\ref{fig1} (d) shows two distinct VHS around the Fermi energy with a finite gap at the CNP. In contrast to the gapless dispersion at CNP of Fig.~\ref{fig1} (b), the finite gap throughout the BZ in presence of $\Delta=12$~meV makes the system insulating (see Fig.~\ref{fig1} (c)). Similarly, we find that the application of a finite electric field in TDBG with $\theta = 1.33^\circ$ leads to an increased energy separation between the VHS of the flat bands near the CNP in Fig.~\ref{fig1} (h).

Clearly the external perturbations such as twist angle and vertical electric field can mediate the metal insulator transition in TDBG, and can tune the location of the VHS as well. This tunability of the electronic spectrum will also translate into the tunability of the collective charge mode or plasmons in \moire systems with VHS. Before demonstrating this, we explore the fundamental physics of the interband and intraband plasmons in \moire systems.

%%%%%%%%%%%%%%%%%%%%%%%%%%%%%%%%%%%%%%%%%%%%%%%%%%%%%%%%%%%%%%%%%%%%%%%%%
\section{Gapped and slow interband plasmons}\label{section-III}
\noindent The combination of electronic flat bands and VHS near the CNP in moir\'{e} superlattices motivates several intriguing questions about their plasmon modes. For example, why do the intraband plasmon modes in 2D metallic moir\'{e} systems deviate from the characteristic $\omega_p \propto \sqrt{q}$  dispersion and become `flat' or non-dispersing? What are the conditions needed for the existence of interband plasmon modes? Is there any connection of the plasmon modes to the VHS in electronic states? 

To address such questions, we calculate the plasmon dispersion of TDBG by evaluating the frequency and momentum dependent longitudinal dielectric function $\varepsilon(\bm{q},\omega)$. Starting from the single-particle Hamiltonian, $H(\bm k)$, we s compute $\varepsilon(\bm{q},\omega)$ by numerically evaluating the non-interacting density-density response function (see Sec.~A2 of SI \cite{Note1} for details). The interacting density-density response function is then obtained within the time dependent Hartree approximation or the random phase approximation (RPA)~\cite{giuliani2005quantum}. Within RPA, the dielectric function is specified by
\begin{eqnarray}\label{dielectric_fn_RPA}
	\varepsilon^{\rm RPA}(\bm{q},\omega)= 1- V_q\Pi(\bm{q},\omega)~.
\end{eqnarray}
Here, $V_q=2\pi e^2/({4\pi\kappa\epsilon_0{q}})$ is Fourier transform of the Coulomb potential in 2D, with $\epsilon_0$ and $\kappa$ denoting the vacuum permittivity and the static dielectric constant of the background substrate, respectively. The collective charge modes are specified by the zeros of the real part of $\varepsilon^{\rm RPA}(\bm{q},\omega)$. 
In Eq.~\eqref{dielectric_fn_RPA}, the noninteracting density-density response function or the `Lindhard' function, $\Pi({\bm q},\omega)$,  can be expressed as ~\cite{giuliani2005quantum,Levitov2019,Dutta_C_2022},
\be \label{equ_Pi}
\Pi(\bm{q},\omega)= {g}\sum_{\bm{k}}\sum_{m,n}\frac{(f_{n,\bm{k+q}}-f_{m,\bm{k}})F^{mn}_{\bm{k,k+q}}}{E_{n,\bm{k+q}}-E_{m,\bm{k}}-\omega-i0}~.
\ee
The prefactor $g$ accounts for the degeneracy of the states, which can arise from the two spin and two valley degrees of freedom~\cite{Levitov2019}.
%, similar to that in graphene and in TBG
The summation runs over the ${\bm k}$-points of the \moire BZ, the indices $m$, $n$ specify the bands and $f_{m,\bm{k}}$ denotes Fermi-Dirac distribution function. $F^{mn}_{\bm{k},\bm{k+q}}=|\langle n,\bm{k+q}|m, \bm{k}\rangle|^2$ describes the band overlap function between the cell periodic part of the Bloch eigenstates at momentum $\bm{k}$ and $\bm{k+q}$. See Sec.~A2 and Sec.~B of the SI~\cite{Note1} for details of the numerical calculation of the dielectric function. 
%We set $\hbar=1$, throughout the calculation.
%
Experimentally, plasmon modes are probed via the `scattering type near field optical microscopy' which does the nano-imaging of plasmonic excitations in the real space or through the inelastic `electron energy loss spectroscopy' (EELS)~\cite{Fei2012, DNBasov2021, Chen2012,EELS2014}.
%Experimentally, plasmon modes are usually probed via the inelastic EELS, which probes the loss function spectrum~\cite{EELS2014}. 
The plasmon modes appear as peaks in the loss function, ${\cal L}({\bm q},\omega)$, spectrum that is related to the dielectric function via
\begin{equation}
{\cal L}(\bm{q},\omega) \approx  -{\rm Im}\left[\frac{1}{\varepsilon^{\rm RPA}(\bm{q},\omega)}\right].
\label{EELS}
\end{equation}

\begin{figure*}[t!]
 	\includegraphics[width=0.98\linewidth]{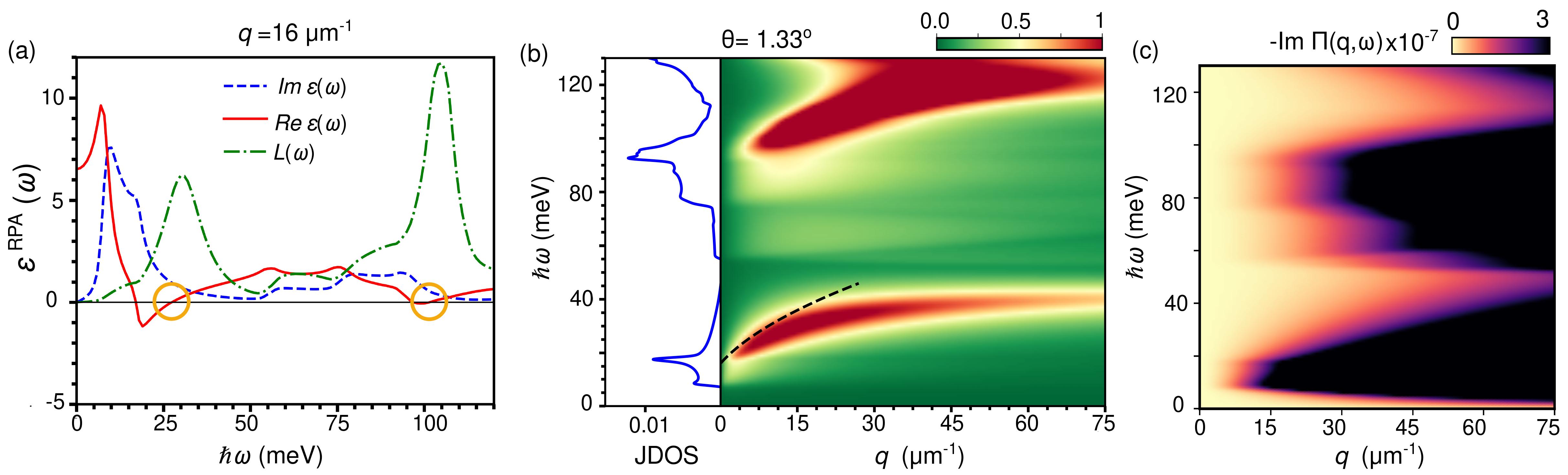}
 	\caption{(a) The dielectric function and (b) the color map of the loss function spectrum, ${\cal L}(\bm{q},\omega)$, for TDBG at twist angle $\theta=1.33^{\circ}$. The black dashed line represents the interband plasmon dispersion as described in Eq.~\eqref{equ_disp_inter}. Note that the interband plasmon peaks at $\sim 16$, and $\sim 95$ meV is aligned with the peaks of JDOS which arise from the transition between the VHS in the DOS as marked in Fig.~\ref{fig1}(e). % Panel.%(c) shows the $\bm{k}$ integrated overlap term in Eq.~\eqref{equ_4} between the 1st valence and 1st conduction bands labed as 1, 2 [see Fig.~\ref{fig1}(d)]. It shows the $q^2$ dependence at small $q$. 
 	(c) Distribution of imaginary part of Lindhard function, $-\mathrm{Im}\Pi(\bm{q},\omega)$ [in units of \AA$^{-2}\mathrm{meV}^{-1}$] in the $q-\omega$ plane for $\theta=1.33^{\circ}$. The purple colored regions represent the electron-hole continuum region  where the collective plasmon modes can decays into electron-hole pairs via Landau damping.
 	}
 	\label{fig2}
\end{figure*}%

We present the calculated $\varepsilon({\bm{q},\omega})$ as a function of $\omega$ in  Fig.~\ref{fig2}(a), for insulating TDBG with $\theta=1.33^\circ$. Here, we have chosen the chemical potential to lie in the CNP gap ($\mu=0$~meV), and a small $q$ value ($\sim16~\mu m^{-1}$) along the $\Gamma-K$ direction. Two distinct zeros in the real part of the dielectric function, marked in orange circles in Fig.~\ref{fig2}(a), are located at $\sim16$ meV, and $\sim110$ meV. Both zeros of the dielectric function are accompanied by peaks in the loss function (green line in Fig.~\ref{fig2} (a)), and have negligible imaginary parts, indicating that these are long lived plasmon modes~\cite{Agarwal_L_2014,Agarwal2015}. Since there are no free carriers at the Fermi energy (see Fig.~\ref{fig1} (d)), both of these plasmon modes arise solely from interband correlations. To probe the features of these interband plasmon modes, we present the distribution of the loss spectrum in the $\omega$ and $q$ (in the $\Gamma-K$ direction) plane, along with the calculated JDOS in Fig.~\ref{fig2}(b). More interestingly, we find that the plasmon gap at $q=0$ is equal to the energy value corresponding to the peak in the JDOS, shown in the left panel of Fig.~\ref{fig2}(b). 

Physically, the  interband plasmon mode is excited by a time dependent electric field, which initially originates either from  internal charge density fluctuations or is applied externally. This dynamical electric field leads to interband transitions predominantly for frequencies corresponding to the JDOS peak simply because of the presence of a large number of transition states. This builds up oscillating charge densities in the system, which further induce a dynamical electric field in the system.  This cycle becomes self sustaining in a resonant condition, which manifests as interband plasmon modes.

The intense plasmon modes seen in the distribution of the loss function in  Fig.~\ref{fig2}(b) clearly show that the interband plasmon modes in TDBG are reasonably long lived. To explicitly confirm that the interband plasmon modes are not Landau damped by electron-hole  excitations, we show the boundary of the single particle excitations in Fig.~\ref{fig2}(c) via the color plot of the imaginary part of the noninteracting density-density response function~\cite{giuliani2005quantum}. It is clear from the plot that the interband plasmon modes predominantly lie outside the electron-hole continuum. The plasmon modes comes close to the electron-hole continuum only for $q > 60~\mu m^{-1}$. The partial damping of the  plasmon mode at sufficiently higher momentum transfer is also  corroborated through the loss in the intensity of the loss function  spectrum. The next interband plasmon peak in the loss spectrum at $\approx 100$~meV originates from the VHS in the JDOS arising from the transition between the flat bands and the higher \moire bands of TDBG.  The color map of ${\cal L}(\bm{q},\omega)$ for $\bm{q}$ along the $\Gamma-M$ direction also shows similar characteristics, as shown in Sec.~C of SI~\cite{Note1}. 
Another interesting feature of the plasmons in \moire superlattices including TDBG, is their extremely `flat' dispersion for large $q$ values [see Fig.~\ref{fig2} (b)]. Similar interband plasmons with flat dispersion have also been predicted in small angle TBG~\cite{Stauber2016,FrankHLKoppen2021, Shengjun2021, marco_tbg20}. However, the origin of the flat plasmon dispersion at large $q$, is not well understood. In addition, fundamental physics questions related to the origin of the interband plasmon modes and the criterion of their existence are still unanswered. We focus on some of these questions below.  

\subsection{\textbf{Origin and dispersion of the interband plasmons}}
\noindent To understand i) the connection of the interband plasmon modes to VHS, and ii) the origin of their flat dispersion, we start from the Lindhard function in Eq.~\eqref{equ_Pi}. The interband part of the density-density response function can be expressed as~\cite{Levitov2019}, 
\bea
\Pi_{\rm inter}(\bm{q},\omega)= 2g\sum_{\bm{k}}\sum_{m,n}^{n\ne m}\frac{f_{m,\bm{k}}F^{mn}_{\bm{k,k+q}} (E_{n,\bm{k+q}}-E_{m,\bm{k}})}{(\omega+i0)^2-(E_{n,\bm{k+q}}-E_{m,\bm{k}})^2}.\nn
\\
\label{equ_Pi_inter}
\eea
Focusing on the lowest plasmon mode in Fig~\ref{fig2} (b), lets work with the pair of \moire flat bands ($m$, $n\in{}$1st conduction/valance band, labeled as 1/2) near the CNP, and the corresponding VHS peaks in the DOS (see Fig.~\ref{fig1}(e)). Using the flat nature of the bands, we can express the energy difference between them as $E_{n,\bm{k}}-E_{m,\bm{k}}=\Delta_0+\delta E_{\bm{k}}$, where $\Delta_0$ is the constant energy difference between the two lowest VHS peaks, and the momentum dependent gap variation $\delta E_{\bm{k}}$ is relatively small. For an insightful, but approximate estimation of the plasmon dispersion, we neglect the $\delta E_{\bm{k}}$  term in the Lindhard function. In fact, this approximation should also work in other systems without flat bands, as the VHS points in the JDOS carry the maximum contribution in the BZ momentum sum in Eq.~\eqref{equ_Pi_inter}. 
\begin{figure*}[t!]
  \includegraphics[width =0.95\linewidth]{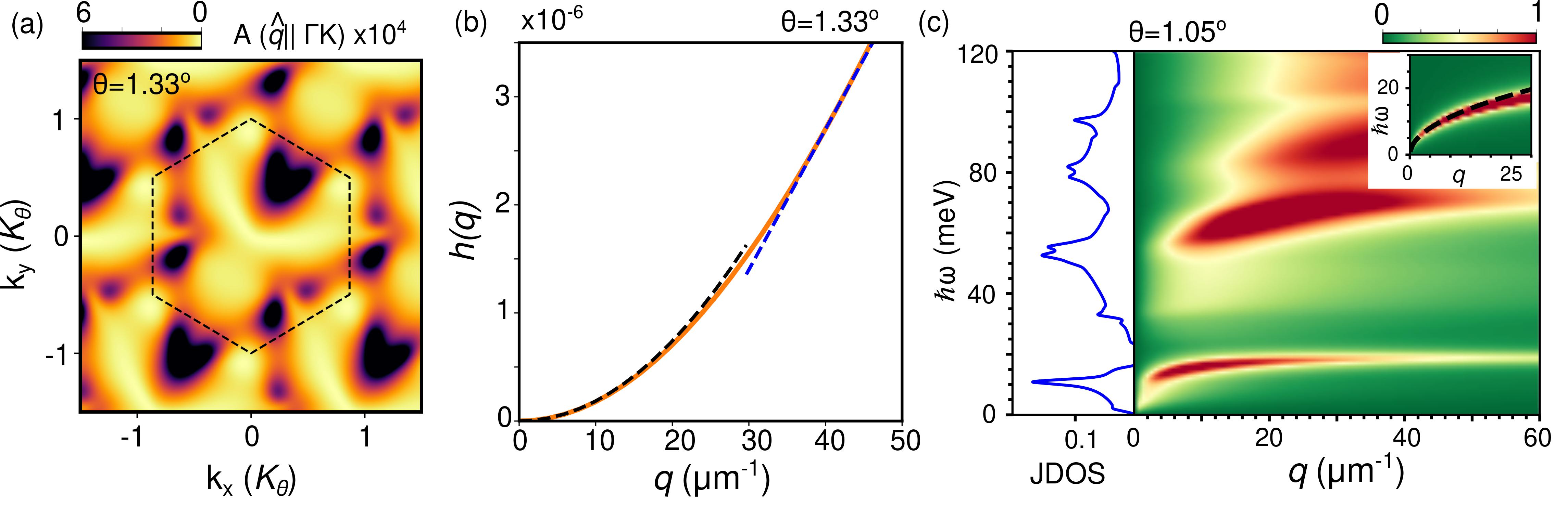}	
	\caption{(a) The distribution of of  A($\hat{q}\parallel\Gamma K$) over $k_x-k_y$ planes (where $k_x$ and $k_y$ are normalized with $K_\theta=8 \pi/(3a) \sin (\theta/2)$ and  $a=2.46~\AA$) for the pair of flat bands near $E_F$ at $\theta=1.33^\circ$. (b) Variation of $h(q)$ with $q$ as defined in Eq.~\eqref{formhq}. Here black dashed line shows quadratic fit up to $q\approx 30~{\mu m}^{-1}$, and blue dashed line shows the linear fit. (c) Color map of $L(\bm{q},\omega)$ over $q$ and $\omega$ space for TDBG near \textit{magic} angle $\theta=1.05^{\circ}$.  The inset shows the calculated loss function spectrum by considering only the intraband contribution in Eq.~\eqref{equ_Pi}  for the flat bands at CNP. The black dashed line in the inset represents the intraband dispersion as described in Eq.~\eqref{intraband}. }
	\label{fig3}
\end{figure*}
With this simplification, the dispersion of the lowest interband plasmons mode for the insulating \moire systems, can now be obtained from $\varepsilon^{\rm RPA}=0$. For 2D systems, this yields  
\be
1- \frac{ge^2}{\kappa\epsilon_0 q}\frac{\Delta_0 }{\omega^2-(\Delta_0)^2} h({\bm q})= 0~, 
\ee
where we have defined, 
\be 
h({\bm q})\equiv \sum_{\bm{k}}\sum_{(m\ne n)\in{1,2}}f_{m,\bm{k}} F^{mn}_{\bm{k, k+q}}~. \label{formhq}
\ee
This gives the interband plasmon dispersion in 2D to be, 
\bea
\omega_{\rm inter}({\bm q}) \approx \Delta_0 \left(1+ \frac{ge^2}{\kappa\epsilon_0 \Delta_0} \frac{h({\bm q})}{q}
\right)^{1/2}~.
\label{omega_inter}
\eea
This analytical model for the dispersion of the interband plasmon mode is one of the most important results of this work.  Eq.~\eqref{omega_inter} establishes the following two criterion (necessary condition) for the existence of interband plasmon modes: i) the presence of a VHS peak in the electronic JDOS, and ii) the interband correlation being finite for those pair of bands which contribute dominantly to the JDOS peak. In fact, the gap of the plasmon mode is specified by the energy of the JDOS peak, while the interband correlations determine its dispersion.

\subsection{\textbf{Gapped interband plasmons in the $q\to 0$ limit} }
\noindent In the long wavelength ($q \to 0$) limit, we have $ h({\bm q})\approx Aq^2$, where, $A(\hat{\bm q})= \sum_{k,m\ne n}f_{m,\bm{k}} |\hat{\bm q}\cdot {\cal R}^{nm}_{\bm k}|^2$ depends on the direction of ${\bm q}$ and the interband Berry connection, ${\cal R}^{nm}_{\bm k} \equiv \langle n,\bm{k}|\nabla_{\bm{k}}|m,\bm{k}\rangle$ of the pair of bands involved. The distribution of $A(\hat{\bm q})$ over the 2D BZ for momentum transfer $q$ along the $\Gamma-K$ direction is presented in Fig.~\ref{fig3}(a) for $\theta=1.33^\circ$.
This gives a {\it universal} analytical form for the long wavelength interband plasmon dispersion, 
\bea
\omega_{\rm inter} ({\bm q}\to 0) \approx \Delta_0 \left(1+ \frac{ge^2 A}{\kappa \epsilon_0\Delta_0}q\right)^{1/2}~.
\label{equ_disp_inter}
\eea 
This highlights i) the universal nature of the 2D interband plasmon dispersion in the long wavelength limit, and ii) the interband plasmon mode exists only in presence of a finite interband Berry connection for those pair of bands that contribute dominantly to the JDOS peak. We have calculated Eq.~\eqref{equ_disp_inter} for the pair of flat bands near CNP ($\Delta_0=$16~meV) at $\theta=$ $1.33^{\circ}$, and shown in Fig.~\ref{fig2} (b) to compare with the numerically calculated lowest interband plasmon dispersion. From the quadratic fit of $h(\bm{q})$, we estimate $A=0.18$, given $q$ in $\mu m^{-1}$ as shown in Fig.~\ref{fig3} (b). Furthermore, we have also checked Eq.~\eqref{equ_disp_inter} for two other examples of graphene, and in a toy model for bilayer graphene with a large effective mass, which also supports interband plasmons (see Sec.~E and Sec.~F of the SI~\cite{Note1} for details). The small $q$ expansion of Eq.~\eqref{equ_disp_inter} reduces to a linear dispersion very close to the origin [see Fig.~\ref{fig2} (b)]~\cite{FrankHLKoppen2021,Zhang2017}. 
Therefore, the observation of interband plasmon highlights the existence of a finite interband Berry connection for those pair of bands which contribute dominantly to the JDOS peak.

\subsection{\textbf{Flat interband plasmons in the large $q$ limit}}
\noindent In contrast to the universal nature of the long wavelength interband plasmon dispersion, the large $q$ limit of the interband plasmon dispersion is determined by the {\it non-universal} factor $h({\bm q})/q$ in Eq.~\eqref{omega_inter}. This nonuniversal factor gives rise to the flat interband plasmon dispersion in \moire superlattices. 
%To show this explicitly, we present the calculated band overlap function $h(\bm{q})$ for the two flat bands near the CNP gap in Fig.~\ref{fig3} (b). As the lowest interband plasmon mode arises from the VHS in the JDOS for the pair of flat bands, the overlap function of these two bands in $h({\bm q})$, primarily dictates the long wavelength dispersion of the interband plasmon mode. 
%
The numerically calculated $h({\bm q})$, for the two flat bands in an insulating TDBG from the continuum Hamiltonian, is shown in Fig.~\ref{fig3} (b) for $q$ along the $\Gamma-K$ direction. For small $q$ values, the numerical curve follows the quadratic relation as shown by black dashed line. %$\sim Aq^2$ with $A= 0.18$. 
More interestingly, Fig.~\ref{fig3}(b) clearly shows that for $q > 20~\mu m^{-1}$, the $h({\bm q})$ curve starts to deviate from the $q^2$ behavior. 
In fact, we find that for $q > 35~\mu m^{-1}$, it becomes linear (indicated by the blue dashed line). 
In this linear regime, the $h({\bm q})/q$ term in Eq.~\eqref{omega_inter} becomes independent of $q$, making the interband plasmon nondispersive or flat. 
Owing to the very small group velocity,
%of the flat band dispersion, 
these flat plasmons are also referred to as `slow' plasmons~\cite{SJLouie2020}. 

To highlight that the interband plasmon modes need not always have a flat dispersion, we explicitly calculate the interband plasmon dispersion of the monolayer graphene and bilayer graphene model with a large effective mass, in Sec.~E and Sec.~F of the SI \cite{Note1}, respectively. 
In both cases, we find that the interband plasmon mode indeed begins from the energy of the VHS in the JDOS. However, we find that, in contrast to being flat in the large $q$ limit, the interband plasmon mode disperses linearly in the monolayer graphene and massive bilayer graphene model. This reaffirms the nonuniversal and  material dependent aspect of the large $q$ limit of the interband plasmon mode.   

\section{Flat intraband plasmon in metallic \moire systems}\label{section-IV}
\noindent In addition to the interband plasmons arising from interband correlations, metallic moire systems also host an intraband plasmon mode arising from density fluctuations around the Fermi surface. The long wavelength (or $q\to 0$) limit of the intraband plasmon dispersion can be computed from charge stiffness or Drude weight ($D$)~\cite{marco_tbg20,Sachdeva_P_2015}, and it is given by %(see Sec.~B of SI\cite{Note1})
\begin{eqnarray}
	\omega_{\rm intra}=\sqrt{\frac{D}{2\pi\kappa\epsilon_0}q}~.
	\label{intraband}
\end{eqnarray}  
The undoped TDBG shows metallic behavior at the CNP for the \textit{magic} angle $\theta=1.05^{\circ}$ [see Fig.~\ref{fig1} (a)]. We present the color plot of the loss function for $\theta=1.05^{\circ}$ TDBG in Fig.~\ref{fig3} (c). The inset shows only the intraband contribution to the loss function, with the black dashed line representing the intraband plasmon dispersion relation of Eq.~\eqref{intraband}. 
The Drude weight is numerically calculated to be $D\approx$13.5~meV in scale of $e^2/{\hbar}^2$ (see Sec.~B of the SI~\cite{Note1} for details).  This intraband mode is also outside the electron-hole continuum and is long lived, as shown in SI~\cite{Note1}.  
Clearly, the low energy intraband plasmon dispersion in TDBG deviates significantly from expected $\sqrt{q}$ behavior $q$ as shown in Fig.~\ref{fig3}(c).
Similarly to the case of the interband plasmon mode, with increasing $q$, the intraband plasmon dispersion in TDBG also becomes flat and nondispersive, giving rise to slow plasmons. This can be clearly seen in Fig.~\ref{fig3} (c), where the intraband plasmon dispersion in the energy window $0$-$20$ meV becomes flat in large $q$ limit.
%, comparable to the half of the moire reciprocal lattice vectors.
This is primarily an influence of the strong interband correlations of the flat bands, which have a JDOS peak at $\sim10$ meV [see Fig.~\ref{fig3} (c)], which drastically modifies the intraband plasmon dispersion. 

To understand the (i) nature of the intraband plasmon dispersion under the influence of large interband transitions, and (ii) the flatness of the dispersion at large $q$, we express the Lindhard function as the sum of two different contributions, $\Pi \approx \Pi_{1} + \Pi_{2}$~\cite{Levitov2019}. These contributions are separated depending on the energy difference $\Delta_{nm}$= $E_{n,\bm{k+q}}-E_{m,\bm{k}}$ being smaller or larger than $\omega$. For example, at the \textit{magic} angle, the TDBG band dispersion has a gap between the flat bands and the higher energy \moire bands. Therefore, to explore the dispersion of the intraband mode, the flat band contribution should be retained in the first term, while the contribution of the higher bands will go in the second term. In the $\Pi_1$ ($\Pi_2$) term, we have $\omega > |\Delta_{nm}|$ ($\omega < \Delta_{nm}$) in the denominator of Eq.~\eqref{equ_Pi}, and we approximate, 
\begin{subequations}
\bea
\Pi_1(\bm{q},\omega) & {\approx} & 2g\sum_{\bm{k}} {\sum_{nm}}^{\prime}f_{m\bm{k}} \frac{ F^{nm}_{\bm{k},\bm{k+q}} }{\omega^2}\left( E_{n,\bm{k+q}} -E_{m,\bm{k}} \right), ~~~~~  \\
\Pi_2(\bm{q},\omega) & {\approx} & -2g\sum_{\bm{k}} {\sum_{nm}}^{\prime\prime}\frac{f_{m\bm{k}}F^{nm}_{\bm{k},\bm{k+q}}}{\left( E_{n,\bm{k+q}} -E_{m,\bm{k}} \right)}~.
\label{Pi_2}        
\eea%
\end{subequations}%
Here, the summation ${\sum}^{\prime}_{nm}$ (${\sum}^{\prime\prime}_{nm}$) runs over the pair of bands such that the eigenvalues satisfy $\omega > |\Delta_{nm}|$ ($\omega < \Delta_{nm}$). This allows us to express the dielectric function as
\bea
\varepsilon^{\mathrm{RPA}}(\bm{q},\omega)= 1 - \frac{B(\bm{q})}{\omega^2}+ C(\bm{q})~,
\eea
where we have defined $B({\bm q})=\omega^2V_q \Pi_1(\bm{q},\omega)~$ and $C({\bm q}) =- V_q\Pi_{2}(\bm{q},\omega)$.
Now, the overall plasmon dispersion can be obtained from the roots of the dielectric function to be, 
\bea
\omega_p^2 ({\bm q})=\frac{B({\bm q})}{1+C({\bm q})}~,
\label{omega_intra}
\eea
which is valid for both small and large $q$ limit.
In the small $q$ limit, the intraband band overlap function $F^{n=m}_{\bm{k},\bm{k+q}}$ dominates compared to the interband overlap function (see Eq.~S19 of SI.~\cite{Note1}). Therefore, $C(\bm{q})$ term in the denominator have vanishingly small contribution in low $q$ limit. So, the plasmon dispersion is predominantly dictated by $B(q)$ (intraband contributions only) and we have 
$\omega_p\approx\sqrt{B(q)} \propto \sqrt{q}$ [see Sec.~D of the SI~\cite{Note1} for details].
%%%%%%%%%%%%%%%%%%%%%%%%%%%%%%
This establishes that, in the long wavelength limit, the intraband plasmon mode in 2D moire systems also disperses as $\sqrt{q}$.

\begin{figure*}[!t]
 	\centering
 	\includegraphics[width=\linewidth]{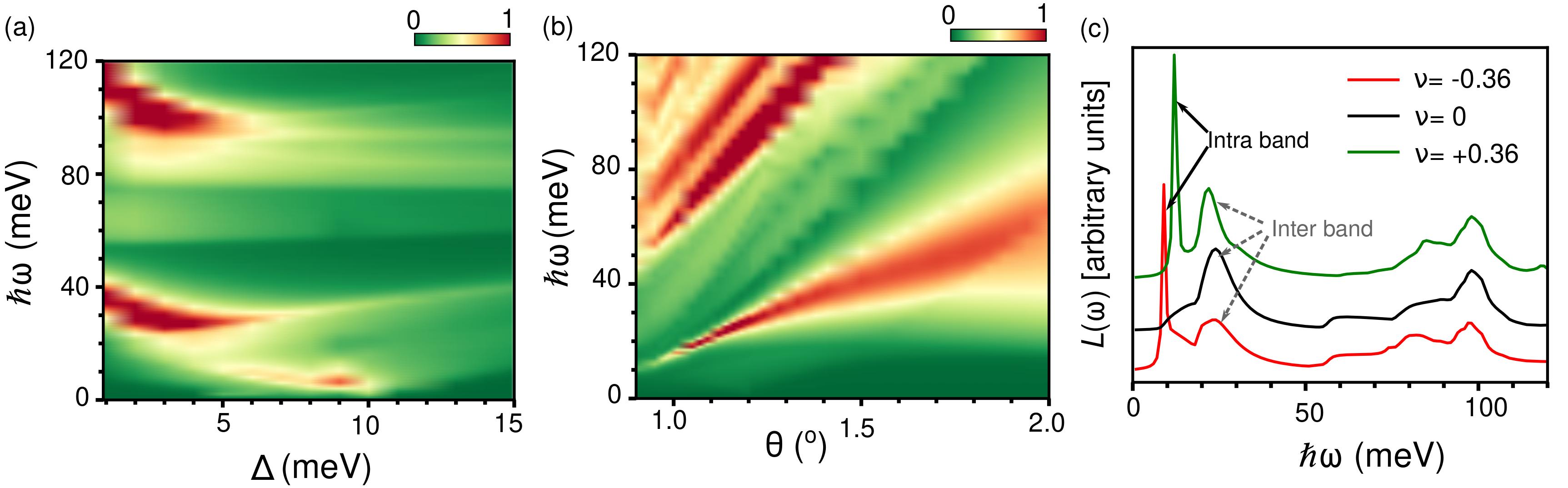}
 	\caption{ (a) Distribution of the loss function with variation of the vertical electric field for  $\theta=1.33^\circ$. The chemical potential is fixed at the CNP. The large tunability of the interband plasmon gap can be clearly seen. The emergence of a new intraband plasmon mode highlights the metal-insulator transition. (b) Variation of the loss function spectrum with changing $\theta$, in absence of external electric field. Clearly, the ladder of plasmon modes persists for a range of twist angles. (c) The loss function spectrum ($\theta=1.33^\circ$ and $\Delta = 0$) for three different occupancies, $\nu=$ -0.36 (red), 0 (black) and +0.36 (green),  highlighting the tunability of the plasmon mode via electron doping. All the figures are plotted at a fixed value of $q=$~5.6 $\mu m^{-1}$.
 	}
\label{fig5} 
\end{figure*}
In the large $q$ limit, comparable to the half of the moire reciprocal lattice vectors, the contribution of $C(q)$ becomes large compared to unity, and it starts influencing the plasmon dispersion.
Consequently, the dispersion of the plasmons is modified to $\omega_p^2 \approx B(q)/C(q)$~\cite{Levitov2019}.  The ratio $B(q)/C(q)$ becomes nearly independent of $q$ for large $q$ in TDBG, and this is what modifies the $\sqrt{q}$ dispersion to become nearly $q$ independent  flat dispersion, supporting very slow plasmons. This clearly highlights the role of strong interband transitions [term $C(q)$] in modifying the intraband plasmon dispersion in the large $q$ limit. 

Having established the universal and the non-universal features of both the intraband and the interband plasmon mode, we now study their tunability.

\section{Tunability of the plasmon modes in TDBG}\label{section-VI}
\noindent One of the experimental advantages of TDBG is the relatively large tunability of the electronic band dispersion as highlighted in Fig.~\ref{fig1}. In this section we demonstrate that the vertical electric field, the twist angle and electronic doping can also be used as knobs to modulate the interband and intraband plasmon dispersion in TDBG. Experimentally, electrostatic gating  via a combination of the top and the back gate can be used to  apply a vertical electric field and to change the electron doping in a controlled way~\cite{Sinha2020,Sinha_B_2022}.

The impact of the vertical electric field on the low energy plasmon modes of small angle TDBG is shown in Fig.~\ref{fig5} (a), for $\theta=1.33^\circ$. With increasing electric field strength, the plasmon peaks in Fig.~\ref{fig5} (a) trace the evolution of the JDOS peaks and mark the metal insulator transition. For $\Delta=0$, the plasmon mode at $\hbar\omega_p \approx 40$ meV originates from the interband transitions between the two flat bands, whereas the correlations between the flat bands and the \moire bands give rise to the plasmon mode around $\sim 100$ meV. On initially increasing the 
electric field, the band gap and the location of the VHS first decreases upto $\Delta_c\sim 7$~meV. This is reflected in the loss function spectrum as a red shift of the plasmon mode. As $\Delta \to  \Delta_c$, the flat bands touch each other at the high symmetry $K$/$K^\prime$ points [see Fig.~S7 (c) in the SI \cite{Note1}] giving rise to a metallic state. The transition to the metallic state is reflected in Fig.~\ref{fig5}(a) as an extra peak (around $\omega \approx 10$ meV) for an  intraband plasmon mode. Further increment in $\Delta > 10$~meV opens up a gap in the electronic spectrum, and consequently the intraband plasmon mode vanishes.  With gradual enhancement of the electric field, the bands become more dispersive, and the VHS peak in the JDOS decreases in strength. This is well captured by the loss function spectrum and is reflected in the diminishing weight of the interband plasmon peaks for $\Delta > 10$ meV. This clearly establishes that the loss function spectrum can also be used to probe the metal insulator transition and the evolution of the JDOS peaks in TDBG and in other \moire systems as well. More interestingly, experimentally accessible electric field values can tune the plasmon frequencies by more than 30\%. 

The variation of the plasmon dispersion on changing the twist angle is captured in Fig.~\ref{fig5} (b). We find that the ladder of plasmon modes in TDBG persists for a wide range of twist angles. TDBG is metallic up to $\theta_c\sim 1.2^{\circ}$ with relatively flat bands that give rise to VHS in the DOS. This is reflected in the sharp low energy intraband plasmon peak for small angle TDBG. When the twist angle becomes larger than a critical value ($\theta_c$), the system becomes insulating, and the sharp intraband plasmon mode evolves into a more dispersive and relatively less intense interband plasmon mode arising from transitions between the flat bands. In addition to this, the energy separation between the bands, and consequently the location of the VHS increases linearly with increasing twist angle. This is also  captured by the loss function in Fig.~\ref{fig5}(b) as the energy of all the interband plasmon modes simply traces the location of the VHS in the JDOS. 

Another control parameter for tuning the low energy plasmon modes in TDBG and other \moire systems is the electron doping ($\nu$). For example, by tuning the chemical potential to lie within the CNP gap or within the \moire gap, we can eliminate the intraband plasmon mode. We show this explicitly in the ${\cal L}(\omega)$ plot for three different electron doping values ($\nu=\pm 0.36$ and 0) in Fig.~\ref{fig5} (c). At $\nu=0$ in Fig.~\ref{fig5} (c), there are only interband plasmon peaks at $\omega_p \approx 30$ and $100$ meV, arising from the VHS in the JDOS for interband transitions. Tuning the chemical potential to be in either the conduction ($\nu = 0.36$) or in the valence ($\nu = -0.36$) band, gives rise to an additional intraband plasmon peak in the loss function spectrum, whose dispersion and intensity can be further tuned by varying the electron doping.

%%%%%%%%%%%%%%%%%%%%%%%%%%%%%%%%%%%%%%%%%%%%%%%%%
\section{Discussion}\label{section-VII}

\noindent We have demonstrated that small angle TDBG hosts several long lived and slow plasmon modes with a flat dispersion in an energy window of 0-100 meV. %The long lived flat plasmons in insulating TDBG are interband plasmons which arise from the interband transitions between the flat bands, and between the flat and \moire bands. 
We showed that the necessary criteria for the existence of the interband  plasmon modes are i) the presence of a VHS in the JDOS, and ii) a finite interband Berry connection between the pair of bands giving rise to the VHS. We find that each of these interband plasmons has a universal gapped dispersion of the form $\omega_p \propto \Delta_0\sqrt{1 + \gamma q}$ in the long wavelength limit, where $\Delta_0$ marks the location of the VHS in the JDOS, and $\gamma$ is a material specific parameter. 
%In the large $q$ limit, the interband plasmons have a nonuniversal material dependent dispersion which becomes flat in \moire systems supporting slow plasmons. 
%
In addition to the interband plasmon modes, metallic TDBG also supports a gapless intraband plasmon mode, which disperses as $\omega_p \propto \sqrt{q}$ as $q\to 0$. The intraband as well as the interband plasmon modes have a nonuniversal dispersion in the large $q$ limit, which becomes flat in \moire superlattices. 
%The flat dispersion of the intraband plasmon mode arises from an intricate interplay of intraband density fluctuations and interband transitions between flat bands and higher \moire bands. %near the charge neutrality point. 

 We have demonstrated that the long lived and flat plasmon modes in TDBG are highly tunable and they can be controlled by the vertical electric field, electron doping and they persist over a wide range twist angle. %We showed that the appearance and disappearance of the intraband plasmon peak in the loss function spectrum are a signature of the metal insulator transition in TDBG. Furthermore, 
 We  have shown that the change of the interband plasmon peak with the twist angle or vertical electric field captures the separation of the flat and \moire bands or the VHS in the JDOS. The low energy,  flat, long lived and tunable terahertz plasmon modes in TDBG offers a unique platform for studying the fundamental aspects and potential application of photon based quantum information processing.

\section*{Acknowledgment}
\noindent  A.C acknowledges Indian Institute of Technology, Kanpur and  Science and Engineering Research Board (SERB) National Postdoctoral Fellowship (PDF/2021/000346), India for funding. We thank Mandar M. Deshmukh and Barun Ghosh for useful comments. We acknowledge the Science and Engineering Research Board (SERB) and the Department of Science and Technology (DST) of the Government of India for financial support. We thank CC-IITK for providing the high performance computing facility. 

%\section{Data availability}
%\noindent All data needed to evaluate the conclusions in the paper are present either in the main manuscript or in the Supplementary Information.

%\section{Code Availability}
%\noindent The codes used to construct the model Hamiltonian and to calculate the density-density response function along with the loss function spectrum are available on reasonable request from the authors. 

%\section{Competing interest}
%The Authors declare no Competing Financial or Non-Financial Interests

%\section{Author contribution}
%A.C., D.D. and A.A designed the project, performed research, analyzed data, and wrote the paper. A.A. supervised the project.

%\bibliographystyle{alpha}
\bibliography{reference.bib}

\end{document}